\documentclass[a4paper,fleqn]{article}
\usepackage{amsmath}
\begin{document}

\title{\textbf{A note on Lax pairs of the Sawada--Kotera equation}}

\author{\textsc{Sergei Sakovich}\bigskip \\
\small Institute of Physics, Academy of Sciences, Minsk, Belarus $\diamond$ saks@tut.by}

\date{}

\maketitle

\begin{abstract}
We prove that the new Lax pair of the Sawada--Kotera equation, discovered recently by Hickman, Hereman, Larue, and G\"{o}kta\c{s}, and the well-known old Lax pair of this equation, considered in the form of zero-curvature representations, are gauge equivalent to each other if and only if the spectral parameter is nonzero, while for zero spectral parameter a non-gauge transformation is required.
\end{abstract}

\section{Introduction}

Recently, the following interesting result was obtained by Hickman, Hereman, Larue, and G\"{o}kta\c{s} \cite{HHLG}. It turned out that the Sawada--Kotera equation \cite{SK,CDG}
\begin{equation}
u_t + 5 u^2 u_x + 5 u_x u_{xx} + 5 u u_{xxx} + u_{xxxxx} = 0 \label{e1}
\end{equation}
possesses two different Lax representations in the operator form
\begin{equation}
L \psi = \lambda \psi , \qquad \psi_t = M \psi , \label{e2}
\end{equation}
where subscripts of the scalar functions $u$ and $\psi$ denote respective derivatives, $L$ and $M$ are linear differential operators expressed in powers of the derivative operator $D_x$, and $\lambda$ is the spectral parameter. The first Lax pair, given by the operators
\begin{align}
L_1 & = D_x^3 + u D_x , \notag \\
M_1 & = 9 D_x^5 + 15 u D_x^3 + 15 u_x D_x^2 + \left( 5 u^2 + 10 u_{xx} \right) D_x , \label{e3}
\end{align}
is well known \cite{DG,FG}. The second Lax pair, given by the operators
\begin{align}
L_2 & = D_x^3 + u D_x + u_x , \notag \\
M_2 & = 9 D_x^5 + 15 u D_x^3 + 30 u_x D_x^2 + \left( 5 u^2 + 25 u_{xx} \right) D_x \notag \\
& \quad + ( 10 u u_x + 10 u_{xxx} ) , \label{e4}
\end{align}
is new, in the sense that it appeared in \cite{HHLG} for the first time in the literature.

Many experts, according to their private communications, noticed that the second Lax pair \eqref{e4} is related to the first Lax pair \eqref{e3} by the transformation
\begin{equation}
L_2 = - {L_1}^{\dagger} , \qquad M_2 = - {M_1}^{\dagger} , \label{bis1}
\end{equation}
where the dagger denotes the Hermitian conjugate. This transformation \eqref{bis1} always turns a Lax pair of an integrable equation into a Lax pair of the same equation, but usually the resulting Lax pair has essentially the same form as the original one (we believe that for this reason no second Lax pair was discovered in \cite{HHLG} for the Kaup--Kupershmidt equation, in particular). Let us note, however, that the Lax pairs \eqref{e3} and \eqref{e4} are different in form. Some other experts, also according to their private communications, noticed that the old Lax pair \eqref{e3} and the new one \eqref{e4} are related to each other by the transformation
\begin{equation}
L_2 = D_x L_1 D_x^{-1}, \qquad M_2 = D_x M_1 D_x^{-1}, \label{bis2}
\end{equation}
which corresponds to the transformation $\psi \mapsto \psi_x$ made in \eqref{e2}. Thus, there exist (at least) two different ways to relate the Lax pairs \eqref{e3} and \eqref{e4} to each other, and we believe that this point deserves further investigation using more general description of Lax pairs than their operator form.

In the present paper, we study these two Lax pairs of the Sawada--Kotera equation \eqref{e1}---the old one, \eqref{e2} with \eqref{e3}, and the new one, \eqref{e2} with \eqref{e4}---in the matrix form
\begin{equation}
\Phi_x = X \Phi , \qquad \Phi_t = T \Phi , \label{e5}
\end{equation}
or, what is the same, in the form of zero-curvature representations (ZCRs)
\begin{equation}
D_t X - D_x T + [ X , T ] = 0 , \label{e6}
\end{equation}
where $\Phi(x,t)$ is a three-component column vector, $X$ and $T$ are $3 \times 3$ matrices, and the square brackets denote the matrix commutator. In Section~\ref{s2}, we show that, for any nonzero value of the spectral parameter, the new Lax pair of the Sawada--Kotera equation and the old one are related to each other by a gauge transformation of ZCRs
\begin{align}
\Phi & \mapsto G \Phi , \qquad \det G \neq 0 , \notag \\
X & \mapsto G X G^{-1} + ( D_x G ) G^{-1} , \notag \\
T & \mapsto G T G^{-1} + ( D_t G ) G^{-1} , \label{e7}
\end{align}
where $G$ is a $3 \times 3$ matrix. In Section~\ref{s3}, we show that, for any value of the spectral parameter including zero, the new Lax pair and the old one are related to each other by a gauge transformation \eqref{e7} combined with a different type of equivalence transformations of ZCRs \eqref{e6}, namely,
\begin{equation}
X \mapsto - \tilde{X} , \qquad T \mapsto - \tilde{T} , \label{e8}
\end{equation}
where the tilde denotes the matrix transpose. Section~\ref{s4} contains concluding remarks.

We use computationally effective techniques, such as the method of gauge-invariant description of ZCRs, developed in \cite{M} and \cite{S1} independently, and the method of cyclic bases of ZCRs \cite{S1,S2}, and follow the terminology and notations adopted in \cite{S2}.

\section{Nonzero spectral parameter} \label{s2}

Introducing the three-component column vector
\begin{equation}
\Phi =
\begin{pmatrix}
\psi \\
\psi_x \\
\psi_{xx}
\end{pmatrix}
, \label{e9}
\end{equation}
we can rewrite the Lax pairs \eqref{e2} with the operators \eqref{e3} and \eqref{e4} in their matrix form \eqref{e5}. The old Lax pair of the Sawada--Kotera equation, determined by the operators \eqref{e3}, corresponds to the ZCR \eqref{e6} with the matrices
\begin{align}
X_1 & =
\begin{pmatrix}
0 & 1 & 0 \\
0 & 0 & 1 \\
\lambda & - u & 0
\end{pmatrix}
, \notag \\[4pt]
T_1 & =
\begin{pmatrix}
6 \lambda u & - u^2 + u_{xx} & 9 \lambda - 3 u_x \\
9 \lambda^2 + 3 \lambda u_x & - 3 \lambda u + v & - u^2 - 2 u_{xx} \\
- \lambda u^2 + \lambda u_{xx} & 9 \lambda^2 + u^3 + 2 u u_{xx} + v_x & - 3 \lambda u - v
\end{pmatrix}
, \label{e10}
\end{align}
where $\lambda$ denotes the spectral parameter, and
\begin{equation}
v = u_{xxx} + u u_x . \label{e11}
\end{equation}
The new Lax pair of the Sawada--Kotera equation, \eqref{e2} with \eqref{e4}, corresponds to the ZCR \eqref{e6} with the matrices
\begin{align}
X_2 & =
\begin{pmatrix}
0 & 1 & 0 \\
0 & 0 & 1 \\
\mu - u_x & - u & 0
\end{pmatrix}
, \notag \\[4pt]
T_2 & =
\begin{pmatrix}
6 \mu u + 3 u u_x + v & - u^2 - 2 u_{xx} & 9 \mu + 3 u_x \\
9 \mu^2 + 3 u u_{xx} + v_x & - 3 \mu u - v & - u^2 + u_{xx} \\
p & q & - 3 \mu u - 3 u u_x
\end{pmatrix}
, \label{e12}
\end{align}
where $\mu$ stands for the spectral parameter, $v$ is given by \eqref{e11}, and
\begin{align}
p & = - \mu u^2 - 2 u^2 u_x + \mu u_{xx} + 2 u_x u_{xx} + 3 u v + v_{xx} , \notag \\
q & = 9 \mu^2 + u^3 - 3 \mu u_x + 2 u u_{xx} . \label{e13}
\end{align}
We have changed the notation for the spectral parameter in \eqref{e12} because no relation between the parameters of \eqref{e10} and \eqref{e12} is assumed initially.

Let us compute the cyclic bases \cite{S1,S2} of the ZCRs \eqref{e6} with the matrices \eqref{e10} and \eqref{e12}, in order to see if there are any obstacles to relate these two ZCRs by a gauge transformation
\begin{equation}
X_2 = G X_1 G^{-1} + ( D_x G ) G^{-1} , \qquad T_2 = G T_1 G^{-1} + ( D_t G ) G^{-1} . \label{e14}
\end{equation}

For the matrix $X_1$ given by \eqref{e10} with a nonzero spectral parameter $\lambda \neq 0$, we find that the cyclic basis is eight-dimensional, consisting of the matrices $C_1 , \nabla_1 C_1 , \nabla_1^2 C_1 , \dotsc , \nabla_1^7 C_1$, where $C_1$ is the characteristic matrix,
\begin{equation}
C_1 = \frac{\partial X_1}{\partial u} =
\begin{pmatrix}
0 & 0 & 0 \\
0 & 0 & 0 \\
0 & - 1 & 0
\end{pmatrix}
, \label{e15}
\end{equation}
and the covariant derivative $\nabla_1$ is defined by the relation $\nabla_1 A = D_x A - [ X_1 , A ]$ with any $3 \times 3$ matrix $A$. The closure equation of the cyclic basis, 
\begin{equation}
\nabla_1^8 C_1 = a_0 C_1 + a_1 \nabla_1 C_1 + a_2 \nabla_1^2 C_1 + \dotsb + a_7 \nabla_1^7 C_1 , \label{e16}
\end{equation}
has the following coefficients in this case:
\begin{align}
a_0 & = 6 u u_x^2 - 6 u_{xx}^2 - 10 u_x v - 2 v_{xxx} + ( 6 u_x u_{xx} + 2 v_{xx} ) v_x / v , \notag \\
a_1 & = 4 u^2 u_x - 34 u_x u_{xx} - 22 u v - 14 v_{xx} \notag \\
& \quad + \left( 27 \lambda^2 + 4 u^3 + 6 u_x^2 + 16 u u_{xx} + 12 v_x \right) v_x / v , \notag \\
a_2 & = - 27 \lambda^2 - 4 u^3 - 10 u_x^2 - 20 u u_{xx} - 12 v_x + 4 u u_x v_x / v , \notag \\
a_3 & = 13 u u_x - 64 v + \left( 9 u^2 + 35 u_{xx} \right) v_x / v , \notag \\
a_4 & = - 9 u^2 - 56 u_{xx} + 21 u_x v_x / v , \qquad a_5 = - 27 u_x + 6 u v_x / v , \notag \\
a_6 & = - 6 u , \qquad a_7 = v_x / v , \label{e17}
\end{align}
where $v$ is given by \eqref{e11}.

For the matrix $X_1$ \eqref{e10} with $\lambda = 0$, we get quite a different situation. In this case, the dimension of the cyclic basis is five, not eight. The closure equation
\begin{equation}
\nabla_1^5 C_1 = a_0 C_1 + a_1 \nabla_1 C_1 + a_2 \nabla_1^2 C_1 + a_3 \nabla_1^3 C_1 + a_4 \nabla_1^4 C_1 \label{e18}
\end{equation}
has the coefficients
\begin{align}
a_0 & = - 2 u u_x - 2 u_{xxx} + 2 u_x u_{xx} / u , \qquad a_1 = - 4 u^2 - 8 u_{xx} + 6 u_x^2 / u , \notag \\
a_2 & = - 6 u_x , \qquad a_3 = - 5 u , \qquad a_4 = u_x / u . \label{e19}
\end{align}

For the matrix $X_2$ \eqref{e12}, which contains $u_x$, the characteristic matrix $C_2$ is computed in the following, more general, way:
\begin{equation}
C_2 = \frac{\partial X_2}{\partial u} - \nabla_2 \left( \frac{\partial X_2}{\partial u_x} \right) =
\begin{pmatrix}
0 & 0 & 0 \\
- 1 & 0 & 0 \\
0 & 0 & 0
\end{pmatrix}
, \label{e20}
\end{equation}
where the covariant derivative $\nabla_2$ is defined by the relation $\nabla_2 A = D_x A - [ X_2 , A ]$ with any $3 \times 3$ matrix $A$. The cyclic basis $C_2 , \nabla_2 C_2 , \nabla_2^2 C_2 , \dotsc , \nabla_2^{n-1} C_2$ for the matrix $X_2$ has the dimension $n = 8$ if $\mu \neq 0$ and $n = 5$ if $\mu = 0$---the same dimensions as for the matrix $X_1$. The coefficients of closure equations in the case of $X_2$ are given by the expressions \eqref{e17}, after the replacement $\lambda^2 \mapsto \mu^2$, for $\mu \neq 0$, and by the expressions \eqref{e19} for $\mu = 0$---the same expressions as for the matrix $X_1$. Taking into account that the dimensions of cyclic bases and the coefficients of closure equations are gauge invariants, we see that the only obstacle for the existence of a gauge transformation \eqref{e14} we have found so far is the condition $\mu^2 = \lambda^2$. This makes sense to try to find the matrix $G$ of \eqref{e14} explicitly.

It is very convenient to make use of the fact that, under the gauge transformation \eqref{e14}, the characteristic matrix and its covariant derivatives transform as tensors \cite{M,S1}, namely,
\begin{equation}
\nabla_2^k C_2 = G \left( \nabla_1^k C_1 \right) G^{-1} , \qquad k = 0 , 1 , 2 , \dotsc . \label{e21}
\end{equation}
Denoting the elements of the matrix $G$ as $g_{ij}$, $i,j = 1,2,3$, we find from the relation
\begin{equation}
C_2 G = G C_1 \label{e22}
\end{equation}
that
\begin{equation}
g_{11} = g_{13} = g_{33} = 0 , \qquad g_{23} = g_{12} . \label{e23}
\end{equation}
Next, we find from the relation
\begin{equation}
( \nabla_2 C_2 ) G = G \nabla_1 C_1 \label{e24}
\end{equation}
that
\begin{equation}
g_{21} = g_{22} = 0 , \qquad g_{32} = - u g_{12} . \label{e25}
\end{equation}
Then, the relation
\begin{equation}
( \nabla_2^2 C_2 ) G = G \nabla_1^2 C_1 \label{e26}
\end{equation}
leads us to
\begin{equation}
g_{31} = \lambda g_{12} , \qquad \mu = \lambda . \label{e27}
\end{equation}
At this point, we can immediately conclude that the conditions $\lambda \neq 0$ and $g_{12} \neq 0$ hold necessarily because $\det G = \lambda g_{12}^3 \neq 0$. Finally, we get
\begin{equation}
D_x g_{12} = D_t g_{12} = 0 \label{e28}
\end{equation}
directly from \eqref{e14}, that is $g_{12} = c$ with any nonzero constant $c$, and obtain
\begin{equation}
G = c
\begin{pmatrix}
0 & 1 & 0 \\
0 & 0 & 1 \\
\lambda & - u & 0
\end{pmatrix}
. \label{e29}
\end{equation}
With the natural choice of $c = 1$ in \eqref{e29}, we have $\det G = \lambda$, and the inverse matrix $G^{-1}$ does not exist for $\lambda = 0$. Of course, one can take $c = \lambda^{-1/3}$ and get $\det G = 1$, but in this case the matrix $G$ does not exist for $\lambda = 0$. As we have already pointed out above, the condition $\lambda \neq 0$ is necessary for the existence of the gauge transformation sought.

Consequently, the two considered ZCRs with the matrices $X$ and $T$ given by \eqref{e10} and \eqref{e12} are related to each other by the gauge transformation \eqref{e14} if and only if $\mu = \lambda \neq 0$, and the corresponding matrix $G$ is given by \eqref{e29}, where one can take $c = 1$ without loss of generality. One can see easily from \eqref{e7}, \eqref{e9} and \eqref{e29} that this gauge transformation corresponds to the transformation \eqref{bis2} between the Lax pairs considered in their operator form. Another way to see this consists in taking into account that $G$ in \eqref{e29} with $c = 1$ is identical to $X_1$ in \eqref{e10}, and therefore we have $\Phi \mapsto G \Phi = D_x \Phi$ in \eqref{e7} owing to \eqref{e5}.

Let us note that it is a new, interesting and quite surprising phenomenon that two ZCRs containing an essential parameter are related to each other by a gauge transformation for all values of the parameter except one value and no gauge transformation exists between those ZCRs for that single value of the parameter.

\section{Arbitrary spectral parameter} \label{s3}

Besides gauge transformations \eqref{e7}, there is a different---quite evident but rarely mentioned in the literature---non-gauge type of equivalence transformations of ZCRs \eqref{e6}, namely, the transformation \eqref{e8}. Let us try to make use of a combination of transformations \eqref{e7} and \eqref{e8} to relate the two ZCRs given by \eqref{e10} and \eqref{e12} to each other.

The problem is to find a matrix $G$ such that
\begin{equation}
X_2 = G X_3 G^{-1} + ( D_x G ) G^{-1} , \qquad T_2 = G T_3 G^{-1} + ( D_t G ) G^{-1} , \label{e30}
\end{equation}
where
\begin{equation}
X_3 = - \tilde{X_1} , \qquad T_3 = - \tilde{T_1} . \label{e31}
\end{equation}
Since the gauge invariants of the cyclic basis in the case of $X_3$ coincide with the ones of $X_1$, we omit their consideration and proceed directly to the analysis of the relations $\nabla_2^k C_2 = G \left( \nabla_3^k C_3 \right) G^{-1}$, $k = 0 , 1 , 2 , \dotsc$, where $\nabla_3$ is defined by $\nabla_3 A = D_x A - [ X_3 , A ]$ for any $3 \times 3$ matrix $A$, and $C_3 = \partial X_3 / \partial u = - \tilde{C_1}$. From the relation $C_2 G = G C_3$, we find for the elements $g_{ij}$ of the matrix $G$ the following: $g_{11} = g_{12} = g_{32} = 0$ and $g_{22} = - g_{13}$. Next, we find from the relation $( \nabla_2 C_2 ) G = G \nabla_3 C_3$ that $g_{21} = g_{23} = 0$ and $g_{33} = - u g_{13}$. Then, the relation $( \nabla_2^2 C_2 ) G = G \nabla_3^2 C_3$ leads us to $g_{31} = g_{13}$ and $\mu = - \lambda$, where $g_{13} \neq 0$ in order to have $\det G \neq 0$. Finally, we get $D_x g_{13} = D_t g_{13} = 0$ directly from \eqref{e30}, set $g_{13} = 1$ without loss of generality, and obtain
\begin{equation}
G =
\begin{pmatrix}
0 & 0 & 1 \\
0 & - 1 & 0 \\
1 & 0 & - u
\end{pmatrix}
. \label{e32}
\end{equation}

Consequently, the two considered ZCRs with the matrices $X$ and $T$ given by \eqref{e10} and \eqref{e12} are related to each other by the combination of transformations \eqref{e30} and \eqref{e31} if and only if $\mu = - \lambda$, and the corresponding matrix $G$ is given by \eqref{e32}. The case of zero spectral parameter is included now. Let us note that we were forced to use the non-gauge transformation \eqref{e8}, which is evidently a counterpart of the transformation \eqref{bis1}, in order to cover the case of zero spectral parameter, because the two studied ZCRs belong to two distinct classes of gauge equivalence if the spectral parameter is zero.

\section{Conclusion} \label{s4}

In this paper, using the method of gauge-invariant description of zero-curvature representations (ZCRs) and the method of cyclic bases of ZCRs, we have shown that the new Lax pair of the Sawada--Kotera equation, discovered recently by Hickman, Hereman, Larue, and G\"{o}kta\c{s}, and the well-known old Lax pair of this equation, considered in the form of ZCRs, are gauge equivalent to each other if and only if the spectral parameter is nonzero, while for zero spectral parameter a non-gauge transformation is required. As a by-product, we have obtained an interesting example of two ZCRs which share the same set of gauge invariants but cannot be related to each other by a gauge transformation.

\section*{Acknowledgments}

The author is grateful to Ziemowit Popowicz, Takayuki Tsuchida, Allan Fordy, and anonymous reviewers for valuable comments.

\end{document}